\documentclass[aps,preprint,showpacs,amsmath,amssymb]{revtex4-1}
\usepackage{graphicx}
\usepackage{dcolumn}
\usepackage{bm}
\usepackage{hyperref}
\usepackage[utf8]{inputenc}
\DeclareUnicodeCharacter{0301}{\'{e}}
\begin{document}

\title{Partial Transpose Moments, Principal Minors and Entanglement Detection}
\author{Mazhar Ali\footnote{Corresponding Author Email: mazharaliawan@yahoo.com or mazhar.ali@iu.edu.sa}}
\affiliation{Department of Electrical Engineering, Faculty of Engineering, Islamic University at Madinah, 
107 Madinah, Saudi Arabia}


\begin{abstract}
Recently, it has been shown that locally randomized measurements can be employed to get partial transpose 
moments of a density matrix [Elben A., {\it et al.} Phys. Rev. Lett. {\bf 125}, 200501 (2020)]. 
Consequently, two general entanglement detection methods were proposed based on partial transpose moments 
of a density matrix [Yu X-D., {\it et al.} Phys. Rev. Lett. {\bf 127}, 060504 (2021)]. 
In this context, a natural question arises that how partial transpose moments are related with entanglement 
and with well known idea of principal minors. In this work, we analytically demonstrate that for 
qubit-qubit quantum systems, partial transpose moments can be expressed as simple functions of principal 
minors. We expect this relation to exist for every bipartite quantum systems. In addition, we have 
extended the idea of PT-moments for tripartite qubit systems and have shown that PT-moments can only detect 
the whole range of being NPT for $GHZ$ and $W$ states mixed with white noise.   
\end{abstract}


\maketitle

\section{Introduction}\label{S-intro}

Quantum entanglement, despite being a peculiar phenomenon, has practical utilization in quantum computation 
and many other important information processing protocols \cite{Hayashi-Book, Wilde-Book}. 
Due to its important and promising role in the upcoming technologies, considerable efforts and resources 
have been allocated towards the characterization and quantification of quantum entanglement 
\cite{Horodecki-RMP-2009, gtreview, Erhard-NRP-2020, Friis-2019}. 

The basic question about entanglement properties of a given quantum state $\rho$, has no clear answer 
except for few special cases. For bipartite systems with Hilbert space $H_A \otimes H_B$, this problem 
has been completely solved for $2 \otimes 2$ and $2 \otimes 3$ quantum systems by {\it Peres-Horodecki} 
criterion (also called PPT criterion) \cite{Peres-PRL77-1996, Horodecki-PLA223-1996}. 
This criterion says that for separable states, the eigenvalues of its partially transposed matrix 
($\rho^{T_A}$ or $\rho^{T_B}$) are all positive (non-negative). Therefore, if any eigenvalue of partially 
transposed matrix is negative then $\rho_{AB}$ is entangled. For Hilbert space 
with dimension larger than 6, there are quantum states having positive partial transpose, nevertheless 
entangled (bound entangled states) \cite{gtreview}. Hence, it is well known that a state being 
PPT (having positive partial transpose) may not be separable. This situation is more complicated for 
multipartite quantum systems, where a given quantum state may be fully separable, bi-separable or 
genuinely entangled. Many efforts have been devoted to study this problem and partial solutions also exist
\cite{Bastian-PRL106-2011,Novo-2013,Hofmann-2014,Guehne-NJP12-2010,Guehne-PLA-2011,Bergmann-2013,Brunner-PRL108-2012, Zhou-QI5-2019, Xu-arxiv}. 
PPT criterion have been extended to multipartite systems and PPT-mixtures were 
defined \cite{Bastian-PRL106-2011}. This method leads to detection of genuine multipartite entanglement. 
If a multipartite state has been detected by this method then it is guaranteed to be genuine entangled 
otherwise it may not be separable. This also means that PPT-entangled 
states are not detected by this technique. Genuine entanglement is different from bipartite entanglement 
in the sense that there are 
multipartite quantum states having negative partial transpose (NPT) for every bipartition but they are not 
genuinely entangled \cite{gtreview}. 

Recent developments in quantum technology are promising the medium scale quantum devices with several 
qubits \cite{Preskill-Q2-2018,Zhong-2020,Zhong-2021,Gong-2021,Wu-2021}. However, the standard quantum tomography may not be feasible to decide 
entanglement properties of a $64$-qubit state (as an example) in experiments \cite{Paris-2004}. Therefore, 
a more direct method to estimate entanglement is desired without bothering the hard task of quantum state 
tomography. For this purpose, locally randomized measurements are proposed to detect and quantify 
entanglement. The measurements are performed on particles in random bases and moments are calculated from 
probabilities. This methods is capable of detecting entanglement \cite{Tran-2015,Tran-2016} and computing 
moments \cite{Enk-2012} for a given density matrix. Locally randomized measurements are also proposed to 
estimate quantum entropies \cite{Elben-PRA-2019,Brydges-2019}, to characterize multipartite entanglement 
\cite{Ketterer-2019, Ketterer-2020,Ketterer-2021,Knips-2020}, 
and to detect bound entanglement \cite{Imai-PRL126-2021}. 
Some authors have studied the verification of PPT condition by using this method 
\cite{Gray-2018, Elben-PRL125-2020, Zhou-2020,Huang-2020}. Recently, realignment-moments were proposed to detect entanglement 
for some quantum states \cite{Zhang-QIP21-2022}.

On the other hand, entanglement can also be verified via principal minors. According to matrix 
theory \cite{Horne-Book}, if a matrix is positive semidefinite, all of its principal minors are non-negative 
and vice versa. Therefore, for an entangled state, the minimum principal minor of its partially transposed 
matrix must be negative. This method has been used to track the trajectory of entanglement under 
decoherence \cite{Huang-PRA76-2007, Ali-JPB42-2009}.
A natural question then arises: What is the relationship of principal minors with partial transpose 
moments (PT-moments)? This question is important not only from an academic point of view but also for 
possible new experimental methods to verify and quantify entanglement. We have answered this question for 
qubit-qubit systems. We have been able to write the PT-moments in terms of principal minors and provide 
clear evidence that both methods are related. Although this might seems intuitive and expected result but 
first it was not known how precisely they are related and second this result might lead to similar 
experimental techniques for detection of entanglement. We expect similar 
relations to exist for all dimensions of bipartite quantum systems. In addition, we have extended the 
idea of PT-moments for tripartite qubit systems. Naturely, we have found the intuitive and expected result 
that PT-moments are able to detect the whole range of NPT states for two important families of quantum 
states. We have shown that for $GHZ$ states and $W$ states mixed with white noise, PT-moments are able to 
detect the NPT region of these states. This is equivalent to say that negativity of tri-partite qubits 
and PT-moments give the same results. We have observed that such extention of PT-moments for multipartite 
system is unable to detect genuine entanglement.

This paper is organized as follows. In section \ref{Sec:PTM}, we briefly review the ideas for partial 
transpose moments and summarize the key results on this problem. We describe principal minors and their 
computation for qubit-qubit systems in section \ref{Sec:PM}. We extend the idea for multipartite systems 
in section \ref{Sec:MPE}. Finally, We conclude the work in section \ref{Sec:conc}. 

\section{Partial transpose Moments for bipartite quantum systems} \label{Sec:PTM}

Let us take $\rho_{AB}$ a quantum state in bipartite system with Hilbert space $H_A \otimes H_B$. 
PPT criterion says that for any separable state $\rho_{AB}^{T_A}$ or $\rho_{AB}^{T_B}$ must be positive 
semidefinite. $T_A$ ($T_B$) denotes transpose of density matrix with respect to subsystem $A$ ($B$). 
In discussions to follow, we will consider only transpose with respect to $A$. 
This criterion is straightforward to check and its violation means the state is entangled. With growing 
number of qubits or qudits, standard method of quantum tomography is both time and resources consuming. 
To address this issue, it was proposed that partial transpose moments (PT-moments), which are feasible 
in experiments, can also detect entanglement. The PT-moments are defined as
\begin{eqnarray}
p_k \equiv Tr \big[ \, (\rho_{AB}^{T_A})^k \, \big] \,, 
\label{Eq:ptm}
\end{eqnarray}
where $k = 1, 2, \ldots d$ with $d$ as dimension of Hilbert space $H_A \otimes H_B$. PT-moments are computed 
from locally random measurements \cite{Zhou-2020}. Recent studies indicate that there is no need to compute 
all $d$ PT-moments, which is itself a daunting task. It is sufficient to compute first few 
PT-moments \cite{Elben-PRL125-2020} and a necessary condition which is called $p_3$-PPT criterion was proposed as 
follows:   
\begin{eqnarray}
\rho_{AB} \in SSS \quad \Rightarrow \quad p_3 \geq p_2^2 \,, 
\end{eqnarray}
where $SSS$ denotes set of separable states. This condition is not sufficient as it can not detect some 
entangled states. This task has been completed in a recent study \cite{Yu-PRL127-2021} and two general 
methods have been proposed to solve the general problem of PT-moments. 

To describe the first general result, let $B_k(P)$ are $(k+1) \times (k+1)$ matrices defined by
\begin{eqnarray}
\big[ \, B_k (P) \, \big]_{ij} \, = \, p_{i+j+1} \, ,
\end{eqnarray}
where $i, j = 0,1, \ldots, k$. These matrices are called Hankel matrices and for $k = 1, 2$, they are 
given as
\begin{eqnarray}
B_1 &=& \left[ 
\begin{array}{cc}
p_1 & p_2 \\ 
p_2 & p_3 
\end{array}
\right] \nonumber\\ 
B_2 &=& \left[ 
\begin{array}{ccc}
p_1 & p_2 & p_3 \\ 
p_2 & p_3 & p_4 \\ 
p_3 & p_4 & p_5 
\end{array}
\right].  
\end{eqnarray}
The first general family of criteria for entanglement detection is following:

{\bf Result 1:} For $p_k = Tr \big[ \, (\rho_{AB}^{T_A})^k \, \big]$ with $k = 1, 2, \ldots, d$, a necessary condition 
for $\rho_{AB}$ being separable is that $B_{\lfloor \frac{d-1}{2} \rfloor} (P) \geq 0$ \cite{Yu-PRL127-2021}. Here 
$\lfloor m \rfloor$ is an integer function. 

It is not difficult to see that $p_1 = 1$ and the first-order criterion $B_1 \geq 0$ implies the condition 
$p_3 \geq p_2^2$. Therefore this result is named as $p_n$-PPT criteria for $n = 3, 5, 7, \ldots$. 

The second general result is a necessary and sufficient condition for PT-moments with order 3 \cite{Yu-PRL127-2021}. We 
only provide the most relevant part which concerns us as follows:

{\bf Result 2:} For $p_k = Tr \big[ \, (\rho_{AB}^{T_A})^k \, \big]$ with $k = 1, 2, 3$, a necessary and 
sufficient condition for $\rho_{AB}$ to be separable is $p_3 \geq \alpha x^3 + (1- \alpha \, x)^3$, where 
$\alpha = \lfloor \frac{1}{p_2} \rfloor$, and $x = \frac{\alpha + \sqrt{\alpha [p_2 (\alpha +1)-1]}}{\alpha (\alpha +1)}$. 
This criterion is called $p_3$-OPPT (optimal PPT) criterion.

\section{Principal minors, PT-moments and entanglement for qubit-qubit systems} 
\label{Sec:PM}

In this section, we study our main aim of relating PT-minors with principal minors for qubit-qubit systems. 
The characterization of entanglement for this dimension of Hilbert space is completely solved. It is known 
that if a $2$-qubit is PPT then it is separable and Peres-Horodecki criterion is necessary and sufficient to 
verify entanglement for this case. It follows that for a separable two-qubit state, its partial transpose 
matrix must have all non-negative eigenvalues, i.e., the matrix 
\begin{eqnarray}
\rho^{T_A} = \left( 
\begin{array}{cccc}
\rho_{11} & \rho_{12} & \rho_{31} & \rho_{32} \\ 
\rho_{21} & \rho_{22} & \rho_{41} & \rho_{42} \\ 
\rho_{13} & \rho_{14} & \rho_{33} & \rho_{34} \\
\rho_{23} & \rho_{24} & \rho_{43} & \rho_{44}
\end{array}
\right).
\label{Eq:abm}
\end{eqnarray}
must be positive semidefinite ($\rho^{T_A} \geq 0$). To find analytical eigenvalues of a $4 \times 4$ matrix 
(with $15$ parameters) is quite complicated and lengthy, if not impossible. It is known that for qubit-qubit 
at most one eigenvalue can be negative \cite{Vidal-PRA65-2002}. An alternative but simpler method to check 
the positivity of a matrix is via principal minor (PM), which is defined as the determinant of a sub-matrix 
obtained by removing some rows and columns from the main matrix. We denote the sub-matrices with symbols 
$\rho(i)$, $\rho(ij)$, $\rho(ijk)$, and $\rho(ijkl)$ and their determinants with symbols 
$[\, \rho(i)\, ]$, $[\, \rho(ij)\, ]$, $[\, \rho(ijk)\, ]$, and $[\, \rho(ijkl)\, ]$. For matrix in 
Eq.~(\ref{Eq:abm}), few sub-matrices are 
\begin{eqnarray}
\rho^{T_A}(4) = \left( 
\begin{array}{c}
\rho_{44} 
\end{array}
\right) \, , \nonumber \\
\rho^{T_A} (23) = \left( 
\begin{array}{cc}
\rho_{22} & \rho_{41} \\ 
\rho_{14} & \rho_{33}  
\end{array}
\right) \, , \nonumber \\
\rho^{T_A} (234) = \left( 
\begin{array}{ccc}
\rho_{22} & \rho_{41} & \rho_{42} \\ 
\rho_{14} & \rho_{33} & \rho_{34} \\ 
\rho_{24} & \rho_{43} & \rho_{44}
\end{array}
\right) \,,
\label{Eq:sm}
\end{eqnarray}
with principal minors 
\begin{equation}
[ \, \rho^{T_A}(4) \, ] = \rho_{44} \, , \nonumber
\end{equation}
\begin{equation}
[ \, \rho^{T_A}(23) \, ] = \rho_{22} \rho_{33} - |\rho_{14}|^2 \,,\nonumber
\end{equation}
\begin{equation}
[ \, \rho^{T_A}(234) \, ] =  \rho_{22} \rho_{33} \rho_{44} - \rho_{22} |\rho_{34}|^2 
- \rho_{33} |\rho_{24}|^2 - \rho_{44} |\rho_{14}|^2 + \rho_{14} \rho_{42} \rho_{43} 
+ \rho_{24} \rho_{34} \rho_{41} \,.
\end{equation}
There are $15$ (``$d^2 -1$'' with $d =4$ in our case) such PMs for matrix in Eq.~(\ref{Eq:abm}). 
Eight of them are confirmed non-negative, i.e., $[\, \rho^{T_A}(1)\, ]$, $[\, \rho^{T_A}(2)\, ]$, 
$[\, \rho^{T_A}(3)\, ]$, $[\, \rho^{T_A}(4)\, ]$, $[\, \rho^{T_A}(12)\, ]$, $[\, \rho^{T_A}(13)\, ]$, 
$[\, \rho^{T_A}(24)\, ]$, and $[\, \rho^{T_A}(34)\, ]$. 
The problem reduces to find the negative PMs among remaining seven, $\{ \, [\, \rho^{T_A}(14)\, ]$, 
$[\, \rho^{T_A}(23)\, ]$, $[\, \rho^{T_A}(123)\, ]$, $[\, \rho^{T_A}(124)\, ]$, $[\, \rho^{T_A}(134)\, ]$, 
$[\, \rho^{T_A}(234)\, ]$, $[\, \rho^{T_A}(1234)\, ] \, \}$. It is easy to see that 
$[\, \rho^{T_A}(12)\, ] = [ \, \rho(12)\, ]$, $[ \, \rho^{T_A}(13)\, ] = [ \, \rho(13)\, ]$, and 
$[\, \rho^{T_A}(34)\, ] = [ \, \rho(34)\, ]$. We also observe that all Bell states and most of entangled 
states have $\rho_{12} = \rho_{13} = \rho_{34} = 0$. Due to these reasons and to have relatively simple 
expressions, we restrict the problem to X-states \cite{Rau-2009}, given as 
\begin{eqnarray}
\rho_X = \left( 
\begin{array}{cccc}
\rho_{11} & 0 & 0 & \rho_{14} \\ 
0 & \rho_{22} & \rho_{23} & 0 \\ 
0 & \rho_{32} & \rho_{33} & 0 \\
\rho_{41} & 0 & 0 & \rho_{44}
\end{array}
\right).
\label{Eq:Xs}
\end{eqnarray}
Eq.~(\ref{Eq:Xs}) with unit trace $\sum_{i=1}^4 \rho_{ii} = 1$, and positivity conditions   
$ \rho_{22} \rho_{33} \geq |\rho_{23}|^2$, and $ \rho_{11} \rho_{44} \geq |\rho_{14}|^2$ comprises quite 
large set of quantum states. All important entangled states are included in X-states and this subset 
captures most of quantum correlations for qubit-qubit systems. X-states are entangled if and only if 
either $\rho_{22} \rho_{33} < |\rho_{14}|^2$ or $\rho_{11} \rho_{44} < |\rho_{23}|^2$. 

We have found that for X-states, we need only two PMs $[\, \rho^{T_A}(14)\, ]$, and $[\, \rho^{T_A}(23)\, ]$ 
to verify the entanglement. All other PMs can be written in terms of them \cite{Ali-JPB42-2009} as 
\begin{equation}
[\, \rho^{T_A} (14)\, ] = \rho_{11} \, \rho_{44} - |\rho_{23}|^2 \, , \nonumber 
\end{equation}
\begin{equation}
[\, \rho^{T_A} (23)\, ] = \rho_{22} \, \rho_{33} - |\rho_{14}|^2 \, , \nonumber 
\end{equation}
\begin{equation}
[\, \rho^{T_A} (123)\, ] = \rho_{11} \, [\, \rho^{T_A}(23)\, ] \, , \nonumber 
\end{equation}
\begin{equation}
[\, \rho^{T_A} (124)\, ] = \rho_{22} \, [ \, \rho^{T_A}(14)\, ] \, , \nonumber 
\end{equation}
\begin{equation}
[\, \rho^{T_A} (134)\, ] = \rho_{33} \, [\, \rho^{T_A}(14)\, ] \,, \nonumber 
\end{equation}
\begin{equation}
[\, \rho^{T_A} (234)\, ] = \rho_{44} \, [\, \rho^{T_A}(23)\, ] \,, \nonumber 
\end{equation}
\begin{equation}
[\, \rho^{T_A} (1234)\, ] = [ \, \rho^{T_A}(14) \, ] \, [\, \rho^{T_A}(23)\, ] \,. 
\end{equation}
As argued above that only one PM can be negative which is directly related with verification and 
quantification of entanglement.

Based on methods described in previous section, we can calculate PT-moments for quantum states $\rho_X$ 
defined in Eq.~(\ref{Eq:Xs}). After some algebra and simplification of terms, we find that all PT-moments 
can be written as a direct and simple function of principal minors. The first six PT-moments in terms of 
principal minors are given as
\begin{eqnarray}
p_1 &=& 1 \,,\nonumber \\
p_2 &=& 1 - 2 \bigg([\, \rho^{T_A} (14) \, ] + [\, \rho^{T_A} (23)\, ] \bigg) - 2 \, g \, h \nonumber \\
p_3 &=& 1 - 3 \, \bigg( g \, [\, \rho^{T_A} (14)\, ] + h \, [\, \rho^{T_A} (23)\, ] \bigg) - 3 \, g \, h 
\nonumber \\
p_4 &=& 1- 4 \, \bigg( g^2 \, [\, \rho^{T_A} (14)\, ] + h^2 \, [\, \rho^{T_A} (23)\, ] \bigg) 
- 4 \, g \, h \nonumber\\&&
+ 2 \bigg([\, \rho^{T_A} (14)\, ]^2 + [\, \rho^{T_A} (23)\, ]^2 \bigg) + 2 \, g^2 \, h^2, \nonumber\\
p_5 &=& 1- 5 \, \bigg( g^3 \, [\, \rho^{T_A} (14)\, ] + h^3 \, [\, \rho^{T_A} (23)\, ] \bigg) 
- 5 \, g \, h\nonumber\\&&
+ 5 \bigg(g \, [\, \rho^{T_A} (14)\, ]^2 + h \, [\, \rho^{T_A} (23)\, ]^2 \bigg) 
+ 5 \, g^2 \, h^2, \nonumber\\ 
p_6 &=&  1 - 6 \, \bigg( g^4 \, [\, \rho^{T_A} (14)\, ] + h^4 \, [\, \rho^{T_A} (23)\, ] \bigg) 
- 6 \, g \, h\nonumber\\&&
+ 9 \bigg(g^2 \, [\, \rho^{T_A} (14)\, ]^2 + h^2 \, [\, \rho^{T_A} (23)\, ]^2 \bigg) 
+ 9 \, g^2 \, h^2 - 2 \, g^3 \, h^3 \, ,
\label{Eq:ptm1}
\end{eqnarray}
where $g = (\rho_{11} + \rho_{44})$ and $h = (\rho_{22} + \rho_{33})$. 

Let us now study this problem for noisy situations, in particular, under entanglement breaking channels (EBC).
It is known that there always exists a channel $\Lambda$ and a pure state $\sigma$ such that any bipartite 
quantum state can be written as $\rho = (I \otimes \Lambda) \, \sigma$ \cite{Werner-2001}. 
In this construction of mixed states, choice of channel and pure states are not unique, 
\begin{equation}
 \rho = (I \otimes \Lambda) \, \sigma = (\Sigma \otimes I) \, \tilde{\sigma}, ,
\end{equation}
where $\Lambda \neq \Sigma$, and $\sigma \neq \tilde{\sigma}$. An important result connecting entanglement 
of mixed states with two channels $\Delta$ and $\Lambda$ acting on second qubit is given as 
\cite{Farias-Science-2009} 
\begin{equation}
 C \big[ \, (I \otimes \Delta) \, \rho \big] = C \big[ \, (I \otimes \Delta \,  \Lambda)
 |\phi^+\rangle\langle \phi^+ | \, \big] \, C[\sigma]\,, 
\end{equation}
where $|\phi^+\rangle = 1/\sqrt{2} (|00\rangle + |11\rangle)$ is maximally entangled Bell state and 
$C(\rho)$ is measure of entanglement called concurrence \cite{Wooters-2001}. 
This equation tells us that entanglement of a mixed state undergoing 
through a noisy channel can be written as a product of concurrence of Bell state under channel 
$\Delta \,  \Lambda$ for second qubit and concurrence of pure state $\sigma$. For entanglement breaking 
channels, it is easy to see that $C \big[ \, (I \otimes \Delta \,  \Lambda)
 |\phi^+\rangle\langle \phi^+ | \, \big]  = 0 $.

We consider a family of mixed entangled states prepared in experiments \cite{Knoll-PRA94-2016}, given as
\begin{equation}
 \rho = (1 - \alpha) \rho_1 + \alpha \, \rho_2 \,,
\end{equation}
where $\rho_1 = |\beta \rangle\langle \beta |$ and $\rho_2 = |0 1\rangle\langle 0 1|$, and 
$|\beta\rangle = \beta |0 0\rangle + \sqrt{1 - |\beta|^2} |1 1 \rangle$. One can find a channel 
$\mathcal{E}_1$ acting on a pure state $|\xi\rangle$ to get $\rho$, i.e.,
\begin{equation}
 \rho = (\mathcal{E}_1 \otimes I) \, |\xi\rangle\langle\xi|\,.
\end{equation}
We can think $\mathcal{E}_1$ as an amplitude damping channel with damping parameter $\eta$ acting on first 
qubit with pure state $|\xi\rangle = \sqrt{\omega} \, |00\rangle + \sqrt{1 - \omega} |11\rangle$. With 
this definition, we can define parameter $\alpha = \eta (1 - \omega)$. The evolution under amplitude damping on 
second qubit gives us the output state 
\begin{equation}
\rho_o = (I \otimes \mathcal{E}_\gamma) \, \rho = (I \otimes \mathcal{E}_\gamma \Lambda) \, \sigma \,.
\end{equation}
It was shown \cite{Knoll-PRA94-2016} that this channel can be written in terms of Kraus operators
\begin{eqnarray}
K_1 &=& \left( \begin{array}{cc}
\sqrt{\frac{\omega + \gamma \, \eta \, (1 - \omega)}{\omega + \eta \, (1 - \omega)}} & 0 \\ 
0 & \sqrt{\frac{\omega (1-\gamma)}{\omega + \gamma \, \eta \, (1 - \omega)}}  
\end{array} \right), \nonumber \\
K_2 &=& \left( \begin{array}{cc}
0 &  0 \\ 
\sqrt{\frac{\eta (1- \gamma )\,(1 - \omega)}{\omega + \eta \, (1 - \omega)}} & 0  
\end{array} \right), \nonumber \\
K_3 &=& \left( \begin{array}{cc}
0 & \sqrt{\gamma} \\ 
0 & 0  
\end{array} \right), \nonumber \\
K_4 &=& \left( \begin{array}{cc}
0 & 0 \\
0 & \sqrt{\frac{\eta \, \gamma \, (1-\gamma) \, (1 - \omega)}{\omega + \gamma \, \eta \, (1 - \omega)}}  
\end{array} \right)\,. 
\end{eqnarray}
Therefore, it is sufficient to check the points at which 
$C \big[ \, (I \otimes \mathcal{E}_\gamma \Lambda) \, |\phi^+\rangle\langle\phi^+| \, \big]$ vanishes, 
because these are the points at which states become PPT. As this evolution keeps the $X$ structure of 
density matrices, therefore all above discussions are valid for noisy processes as well. 
For the example we took, $\rho_{23}(t) = 0$, therefore $[\, \rho^{T_A} (14)\, ]$ is always 
positive. The other principal minor $[\, \rho^{T_A} (23)\, ]$ will be negative as long as 
\begin{eqnarray}
\gamma < \frac{\omega \big[ \, \eta + \omega (1-\eta)}{\eta^2 \, (1-\omega)^2 + \omega (1-\omega) \, \eta}\,.
\label{Eq:ex}
\end{eqnarray}
It was shown \cite{Knoll-PRA94-2016} that concurrence vanished for $\eta = 0.21$, and $\omega = 0.12$ at 
$\gamma \approx 0.65$. This is precisely the value which we get using Eq.(\ref{Eq:ex}) as well.  

%

\section{Principal minors and PT-moments for multipartite systems}
\label{Sec:MPE}

Let us consider three-qubit system to establish the relationship between principal minors and PT-moments. 
As there are $3$ partitions, so one can take partial transpose w.r.t. any of subsystems $A$, $B$, and $C$. 
First, we define the PT-moments for each system as 
\begin{eqnarray}
p^A_k \equiv Tr \big[ \, (\rho_{ABC}^{T_A})^k \, \big] \,, 
\label{Eq:ptm}
\end{eqnarray}
\begin{eqnarray}
p^B_k \equiv Tr \big[ \, (\rho_{ABC}^{T_B})^k \, \big] \,, 
\label{Eq:ptm}
\end{eqnarray}
and 
\begin{eqnarray}
p^C_k \equiv Tr \big[ \, (\rho_{ABC}^{T_C})^k \, \big] \,, 
\label{Eq:ptm}
\end{eqnarray}
where $k = 1, 2, \ldots,d$. We define the tripartite PT-moment as
\begin{equation}
 p_k = (p^A_k \, p^B_k \, p^C_k)^{1/3} \, .
\end{equation}
If we restrict to symmetric states where $p^A_k = p^B_k = p^C_k$, then partial transpose w.r.t any 
partition is sufficient to consider. 

In general, there are $63$ principal minors for the partial transposed 
matrix and some of them would be positive and some negative. The general analysis is too complicated but if 
we restrict ourselves with $X$-states (all other off-diagonal elements are zero 
except $\rho_{18}, \, \rho_{27}, \, \rho_{36}, \, \rho_{45}$), then there are only $4$ principal minors 
which can be negative, namely $[\, \rho^{T_A} (18)\, ]$, $[\, \rho^{T_A} (27)\, ]$, 
$[\, \rho^{T_A} (36)\, ]$, and $[\, \rho^{T_A} (45)\, ]$. The fifth principal minor is equal to the 
product of these four principal minors
\begin{equation}
[\, \rho^{T_A} (12345678)\, ] = [\, \rho^{T_A} (18)\, ] \, [\, \rho^{T_A} (27)\, ] \, 
[\, \rho^{T_A} (36)\, ] \, [\, \rho^{T_A} (45)\, ]\,. 
\end{equation}
These four principal minors are directly related with potential four negative eigenvalues 
of partially transposed matrix $\rho^{T_A}_{ABC}$, respectively.

Let us consider GHZ states mixed with white noise 
\begin{eqnarray}
\rho = \, \alpha \, \frac{\mathcal{I}}{8} + (1 - \alpha) \, |GHZ\rangle\langle GHZ | \, , 
\end{eqnarray}
where $ 0 \leq \alpha \leq 1$ and $|GHZ\rangle = 1/\sqrt{2} (|000\rangle + |111\rangle)$. It is 
known that these states are NPT for $0 \leq \alpha < 0.8$, and genuinely entangled for 
$0 \leq \alpha \leq 0.571$ \cite{Bastian-PRL106-2011}. For $0.571 < \alpha < 0.8$, the states are NPT but 
bi-separable. The states are fully separable (also PPT) for $0.8 \leq \alpha \leq 1$. The possible negative 
eigenvalue for partial transpose matrix is ``$1/8 \, (-4 + 5 \, \alpha)$'' which is negative for 
$0 \leq \alpha < 0.8$. The only negative principal minor is 
\begin{equation}
[\, \rho^{T_A} (45)\, ] = \frac{32 \, \alpha - 15\, \alpha^2 -16}{64}\,, 
\end{equation}
which is negative for $0 \leq \alpha < 0.8$. Finally, $p^A_k = p^B_k = p^C_k$, and first five PT-moments 
for these states are
\begin{eqnarray}
p_1 &=& 1 \,,\nonumber \\
p_2 &=& 1 - \frac{7 \, \alpha}{4} + \frac{7 \, \alpha^2}{8} \nonumber \\
p_3 &=& \frac{1}{64} \, \big[ 16 - 24\, \alpha + 3 \, \alpha^2 + 6 \, \alpha^3 \big] \nonumber \\
p_4 &=& \frac{1}{512} \, \big[ 128 - 448\, \alpha + 624 \, \alpha^2 - 412 \, \alpha^3 
+ 109 \, \alpha^4 \big] \nonumber \\
p_5 &=& \frac{1}{4096} \, \big[256 - 640\, \alpha + 160\, \alpha^2 + 880 \, \alpha^3 
- 955 \, \alpha^4 + 300 \, \alpha^5 \, \big] \,. 
\end{eqnarray}
It turns out that $B_1 < 0$ for $0 \leq \alpha \leq 0.67$ whereas $B_2 < 0$ for $0 \leq \alpha < 0.8$. 
Hence a complete range of being NPT is detected by adding more PT-moments.

As another example, we take $W$-state mixed with white noise
\begin{eqnarray}
\rho_w = \, \beta \, \frac{\mathcal{I}}{8} + (1 - \beta) \, |W\rangle\langle W| \, , 
\end{eqnarray}
where $ 0 \leq \beta \leq 1$ and $|W\rangle = 1/\sqrt{3} (|001\rangle + |010\rangle + |100\rangle)$. It 
can be easily checked that $\rho_w$ are NPT for $0 \leq \beta \leq 0.79$, and genuine entangled for 
$0 \leq \beta \leq 0.521$ \cite{Bastian-PRL106-2011}. The only possible negative principal minor is 
directly related with state being NPT. The $p^A_k = p^B_k = p^C_k$, and first five PT-moments 
for these states are
\begin{eqnarray}
p_1 &=& 1 \,,\nonumber \\
p_2 &=& 1 - \frac{7 \, \beta}{4} + \frac{7 \, \beta^2}{8} \nonumber \\
p_3 &=& \frac{1}{192} \, \big[ 64 - 120\, \beta + 57 \, \beta^2 + 2 \, \beta^3 \big] \nonumber \\
p_4 &=& \frac{1}{41472} \, \big[ 12800 - 44288\, \beta + 59952 \, \beta^2 - 37916 \, \beta^3 
+ 9533 \, \beta^4 \big] \nonumber \\
p_5 &=& \frac{1}{331776} \, \big[45056 - 161280\, \beta + 211840\, \beta^2 - 111920 \, \beta^3 
+ 8565 \, \beta^4 + 7820 \, \beta^5 \, \big] \,. 
\end{eqnarray}
We find that $B_1 < 0$ for $0 \leq \beta \approx 0.629$, whereas $B_2 < 0$ for $0 \leq \beta \leq 0.778$. It 
is evident that $B_3$ would detect the full range of these states as being NPT. 

\section{Discussion and Summary} \label{Sec:conc}

We have studied the relationship between principal minors and PT-moments. Both of them are able to 
detect entanglement for a given quantum state. For two-qubits states, we have 
shown that PT-moments are directly related with principal minors and can be written as their 
simple functions. As both methods lead to same conclusions about a quantum state, therefore it is important 
to study whether it is more economical to detect PT-moments or indirectly via principal minors. 
In addition, we have extended the idea of PT-moments to tripartite systems. We have found that for 
two important families of quantum states, PT-moments can detect a state as being NPT. In addition, 
PT-moments can be written in terms of its principal minors. We have observed that principal minors are also 
directly related with negativity as for many interesting quantum states, both are based on same condition on 
density matrix elements. It is obvious that the way these methods are provided, they can only detect 
entanglement of NPT states. There still remains questions of detecting genuine multipartite entanglement 
via PT-moments or functions of principal minors. We aim to explore these questions in future studies. 

\begin{acknowledgments}
The author is grateful to both referees for their valuable comments.  
\end{acknowledgments}

\end{document}